\documentclass{article}
\usepackage{spconf,amsmath,graphicx,subcaption,multirow,booktabs,hyperref,float,cite,makecell,xcolor}

\title{Polarization Denoising and Demosaicking: Dataset and Baseline Method}
\name{Muhamad Daniel Ariff Bin Abdul Rahman, Yusuke Monno, Masayuki Tanaka, and Masatoshi Okutomi}
\address{Institute of Science Tokyo, Japan}
\begin{document}
%
\maketitle
\begin{abstract}

A division-of-focal-plane (DoFP) polarimeter enables us to acquire images with multiple polarization orientations in one shot and thus it is valuable for many applications using polarimetric information. The image processing pipeline for a DoFP polarimeter entails two crucial tasks: denoising and demosaicking. While polarization demosaicking for a noise-free case has increasingly been studied, the research for the joint task of polarization denoising and demosaicking is scarce due to the lack of a suitable evaluation dataset and a solid baseline method. In this paper, we propose a novel dataset and method for polarization denoising and demosaicking. Our dataset contains 40 real-world scenes and three noise-level conditions, consisting of pairs of noisy mosaic inputs and noise-free full images. Our method takes a denoising-then-demosaicking approach based on well-accepted signal processing components to offer a reproducible method. Experimental results demonstrate that our method exhibits higher image reconstruction performance than other alternative methods, offering a solid baseline. 

\end{abstract}
\section{Introduction}
\label{sec:Intro}
\vspace{-1mm}

Polarization is one of the physical properties of the light describing the strength and orientation of the oscillations of an electromagnetic wave~\cite{goldstein2017polarized}. It has been shown that polarization information is useful for many imaging and vision applications, such as specularity removal~\cite{jospin2019embedded}, reflection removal~\cite{lei2020}, and 3D reconstruction~\cite{cui2017polarimetric}. 

\begin{figure*}
\centerline{\includegraphics[width=\linewidth]{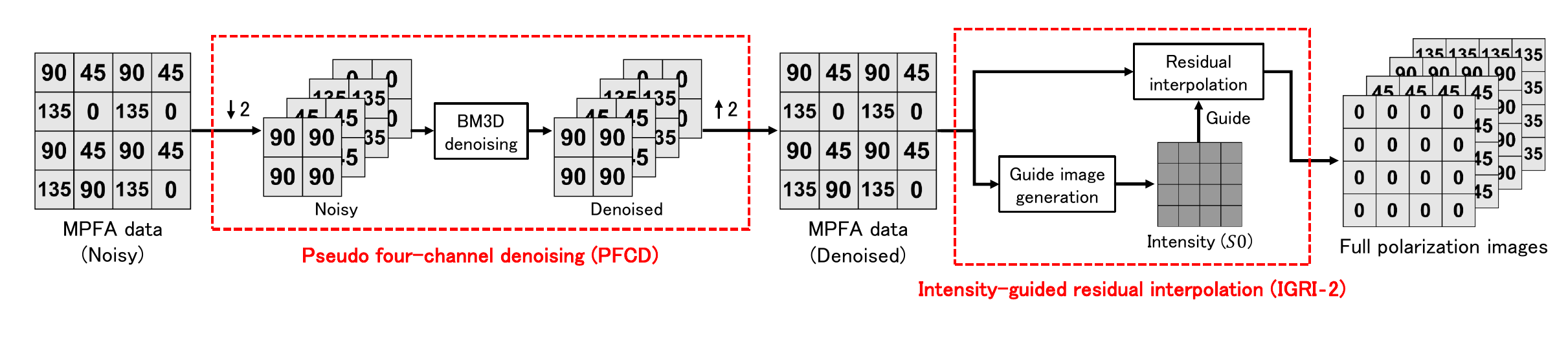}}
\vspace{-6mm}
\caption{The overall flow of our proposed method for MPFA}
\label{fig:OutlineMono}
\end{figure*}

Two representative methods for acquiring polarization images are division-of-time~(DoT) and division-of-focal-plane~(DoFP) methods. The DoT method places a polarizer in front of a camera and sequentially rotates it to obtain a set of images with different polarizer angles~\cite{tyo2006review}. The DoFP method equips a polarization filter array (PFA) on an image sensor, which consists of a mosaic pattern of the pixels with different polarizer angles~\cite{maruyama20183}. Typical examples of Sony's monochrome and color patterns are shown in Figs.~\ref{fig:OutlineMono} and~\ref{fig:OutlineColor}, respectively. While the DoT method is not suitable for dynamic scenes because of the necessity of multiple shots, the DoFP method applies to dynamic scenes owing to its one-shot nature of image acquisition.

The image processing pipeline for a DoFP sensor entails two crucial tasks for high-quality image reconstruction: denoising and demosaicking. Polarization demosaicking is the task of interpolating missing pixel values from raw mosaic PFA data. It has been actively studied in recent years, such as interpolation-based~\cite{liu2020new,lu2024hybrid,morimatsu2021monochrome}, reconstruction-based~\cite{qiu2021linear,wen2021sparse,luo2024learning}, and deep-learning-based~\cite{nguyen2022two,guo2024attention,zheng2024color} methods, induced by the fabrications of Sony's color and monochrome polarization sensors~\cite{maruyama20183}. The datasets provided in~\cite{morimatsu2021monochrome,qiu2021linear} are now becoming standard evaluation platforms. However, these methods and datasets are developed for the demosaicking of noise-free inputs, without explicitly considering denoising. 

The joint task of polarization denoising and demosaicking has also been addressed in some recent studies~\cite{tibbs2018denoising,abubakar2020hybrid,liang2022bm3d,li2020learning,li2023polarized,li2023joint,raffoul2024micropolarizer}. However, there is no consensus on the evaluation platform because there is no publicly available dataset and method for this joint task, to our knowledge. With this in mind, we aim to offer a novel dataset and baseline method for polarization denoising and demosaicking. Our main contributions are as follows.
\begin{itemize}
\vspace{-2mm}
\item We propose a novel real-world dataset containing 40 scenes and three noise-level conditions, consisting of pairs of noisy mosaic inputs and noise-free full ground-truth images as a new evaluation platform.
\vspace{-2mm}
\item We propose a new denoising-then-demosaicking method based on well-accepted signal processing techniques, offering a reproducible and solid baseline method.
\vspace{-2mm}
\item We experimentally demonstrate that our proposed method achieves higher performance than other alternative methods\footnote{Our dataset and source code are available: \url{http://www.ok.sc.e.titech.ac.jp/res/PolarDem/main.html}}.
\end{itemize} 

\vspace{-3mm}
\section{Related Work}
\label{sec:Related}

\subsection{Polarization Demosaicking}
\label{sec:Related_pdm}

Existing polarization demosaicking methods are roughly classified into three categories: interpolation-based, reconstruction-based, and deep-learning-based methods. In this section, we briefly review existing methods by focusing on recent ones and refer to a survey paper~\cite{mihoubi2018survey} for earlier methods.

Interpolation-based methods derive a full polarization image by interpolating missing pixel values in raw mosaic data. Advanced methods adapt well-accepted color demosaicking techniques to polarization demosaicking, such as the exploitations of inter-channel correlations~\cite{liu2020new,lu2024hybrid} and residual interpolation~\cite{morimatsu2021monochrome}. Reconstruction-based methods formulate polarization demosaicking as a linear inverse problem~\cite{qiu2021linear} or a sparse reconstruction problem~\cite{wen2021sparse,luo2024learning} and solve it using an optimization technique. Deep-learning-based methods learn a network that maps the raw mosaic data to a full polarization image in the intensity~\cite{nguyen2022two,guo2024attention} or Stokes-vector~\cite{zheng2024color} domain. While some of these methods include smoothing functionality, such as the incorporation of guided filtering~\cite{liu2020new} and smoothness reqularization~\cite{qiu2021linear}, all of the above methods do not consider denoising explicitly.

Regarding the evaluation for polarization demosaicking, the datasets provided in~\cite{morimatsu2021monochrome,qiu2021linear} are gaining popularity and becoming the standards for evaluation platforms. However, these datasets only provide noise-free inputs and thus non-realistic synthetic noise needs to be added when applied to the evaluation for polarization denoising and demosaicking.

\subsection{Polarization Denoising and Demosaicking}
\label{sec:Related_denoise}

Some methods address the task of polarization denoising and demosaicking. Three approaches can be considered for this task: (i) demosaicking-then-denoisng, (ii) denoising-then-demosaicking, and (iii) joint reconstruction approaches.

The interpolation-based method of~\cite{tibbs2018denoising} takes the first approach and adopts BM3D denoising~\cite{dabov2007image} to the demosaicked images in Stokes-vector domain. The other interpolation-based methods~\cite{abubakar2020hybrid,liang2022bm3d}  take the second approach and apply BM3D or a combined BM3D and K-SVD~\cite{aharon2006k} directly to the raw mosaic data before demosaicking. The deep-learning-based methods of~\cite{li2023polarized,li2023joint} address the task of denoising and demosaicking by learning image reconstruction networks using training images. Although this data-driven manner can be considered as a joint approach, both methods start from initial denoising, leaning toward the denoising-first approach. Recently, a joint method based on wavelet-based Bayesian estimation is also proposed~\cite{raffoul2024micropolarizer}.
Although these methods demonstrate their effectiveness in each paper using synthetic noise or own real data, their source codes and datasets are not publicly available. This motivates us to construct a publicly available real-world dataset and an open-source baseline method for fostering future research.  

\begin{figure*}
\centerline{\includegraphics[width=\linewidth]{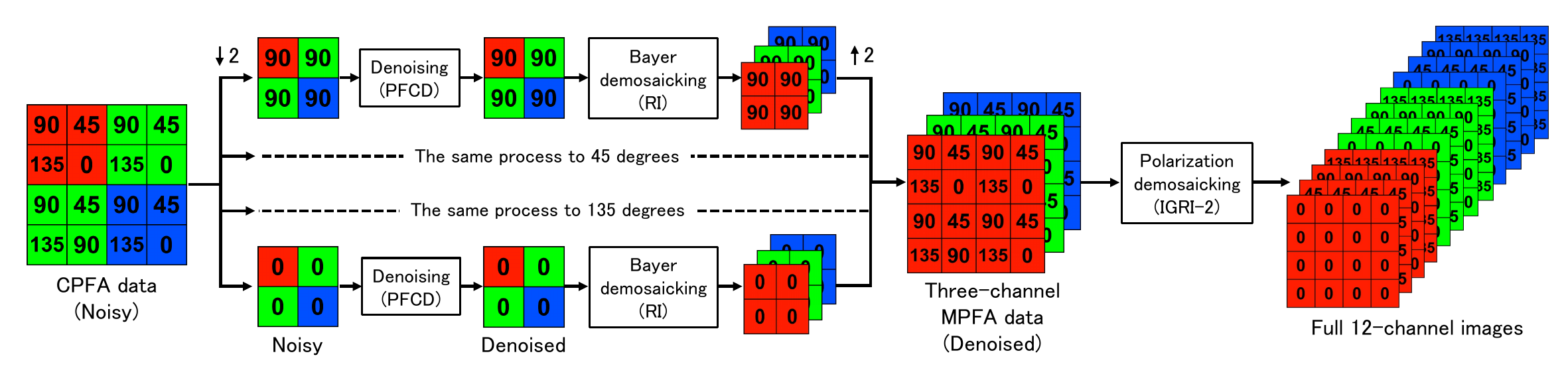}}
\vspace{-5mm}
\caption{The overall flow of our proposed method for CPFA}
\label{fig:OutlineColor}
\end{figure*}

\section{Proposed Method}
\label{sec:Propose} 

\vspace{-1mm}
\subsection{Monochrome PFA}
\label{sec:Monojoint}
\vspace{-1mm}

Figure~\ref{fig:OutlineMono} shows the overview of our method for a monochrome PFA (MPFA) with Sony's sensor pattern~\cite{maruyama20183}. Our method takes a denoising-then-demosaicking approach based on the effective utilization of well-accepted signal-processing-based denoising and demosaicking methods, offering an easy-to-follow and reproducible method as a solid baseline.  

For the denoising stage, we utilize a pseudo four-channel denoising (PFCD) method of~\cite{akiyama2015pseudo}. PFCD was originally developed for the Bayer color pattern~\cite{bayer1976color} and applies the denoising to the pseudo four channels of $(I_{R},I_{G},I_{G},I_{B})$. To exploit PFCD to the MPFA, the pseudo four-channel images of $(I_{0},I_{45},I_{90},I_{135})$ are firstly generated from the raw MPFA data by sub-sampling and re-sampling (denoted as~$\downarrow$2 in Fig.~\ref{fig:OutlineMono}). Then, BM3D denoising~\cite{dabov2007image} is applied in a principal component analysis~(PCA) transformed domain to exploit inter-channel correlations. Specifically, the four-channel pixel values are transformed as $[P_{1},P_{2},P_{3},P_{4}]^T={\bf A}[I_{0},I_{45},I_{90},I_{135}]^T$, where {\bf A} is a 4$\times$4 matrix representing the PCA transform and $P_{i}$ is the value of $i$-th principal component. Accordingly, the noise variances of each channel are derived as $[\sigma^2_{1},\sigma^2_{2},\sigma^2_{3},\sigma^2_{4}]^T={\bf A'}[\sigma^2_{0},\sigma^2_{45},\sigma^2_{90},\sigma^2_{135}]^T$, where ${\bf A'}$ is a matrix whose element is the squared of the corresponding element in ${\bf A}$. After this PCA transformation, BM3D denoising is applied in a channel-by-channel manner using the pair of ($P_{i},\sigma_i$) for $i$-th component. Then, the denoised principal component images are inversely transformed into the four-channel polarization images and rearranged to the original pattern of the MPFA (denoted as $\uparrow$2 in Fig.~\ref{fig:OutlineMono}).

For the demosaicking step, we apply intensity-guided residual interpolation~(IGRI-2)~\cite{morimatsu2021monochrome} to the denoised MPFA data. IGRI-2 first generates an intensity image~($S_0$), which corresponds to the average of four polarization angle images, from the MPFA data in an edge-aware directional manner. Then, RI is performed in a channel-by-channel manner to sub-sampled sparse data of each polarization angle using the generated intensity image as a guide for guided interpolation. By this interpolation process, denoised full polarization images are obtained, from which polarimetric parameters, such as the Stokes parameters $(S_{0},S_{1},S_{2})$ and the angle and degree of polarization ($AoP$ and $DoP$) can be obtained.

\subsection{Color PFA}
\label{sec:Colorjoint}

Figure~\ref{fig:OutlineColor} shows the overall flow of our method for a color PFA~(CPFA) with Sony's sensor pattern~\cite{maruyama20183}. To extend our denoising-then-demosaicking approach to the CPFA, we first generate four Bayer-patterned images corresponding to four polarization angles. Then, PFCD is applied to each Bayer-patterned image, where the pseudo four channels are formed as $(I_{\theta,R},I_{\theta,G},I_{\theta,G},I_{\theta,B})$, where $\theta=\{0,45,90,135\}$ represents the polarization angle. The demosaicking step is separated into two steps: Bayer color demosaicking and polarization demosaicking. Bayer color demosaicking is first applied to the denoised Bayer-patterned images of each polarization angle. Then, those color-demosaicked images are rearranged into three-channel MPFA data. Finally, polarization demosaicking is applied to each MPFA data in a channel-by-channel manner to derive the denoised full 12-channel polarization images. In this paper, we adopt RI~\cite{kiku2016beyond} for Bayer demosaicking and IGRI-2~\cite{morimatsu2021monochrome} for polarization demosaicking.

\section{Proposed Dataset}
\label{sec:Data}

\begin{figure}[t!]
\centerline{\includegraphics[width=\linewidth]{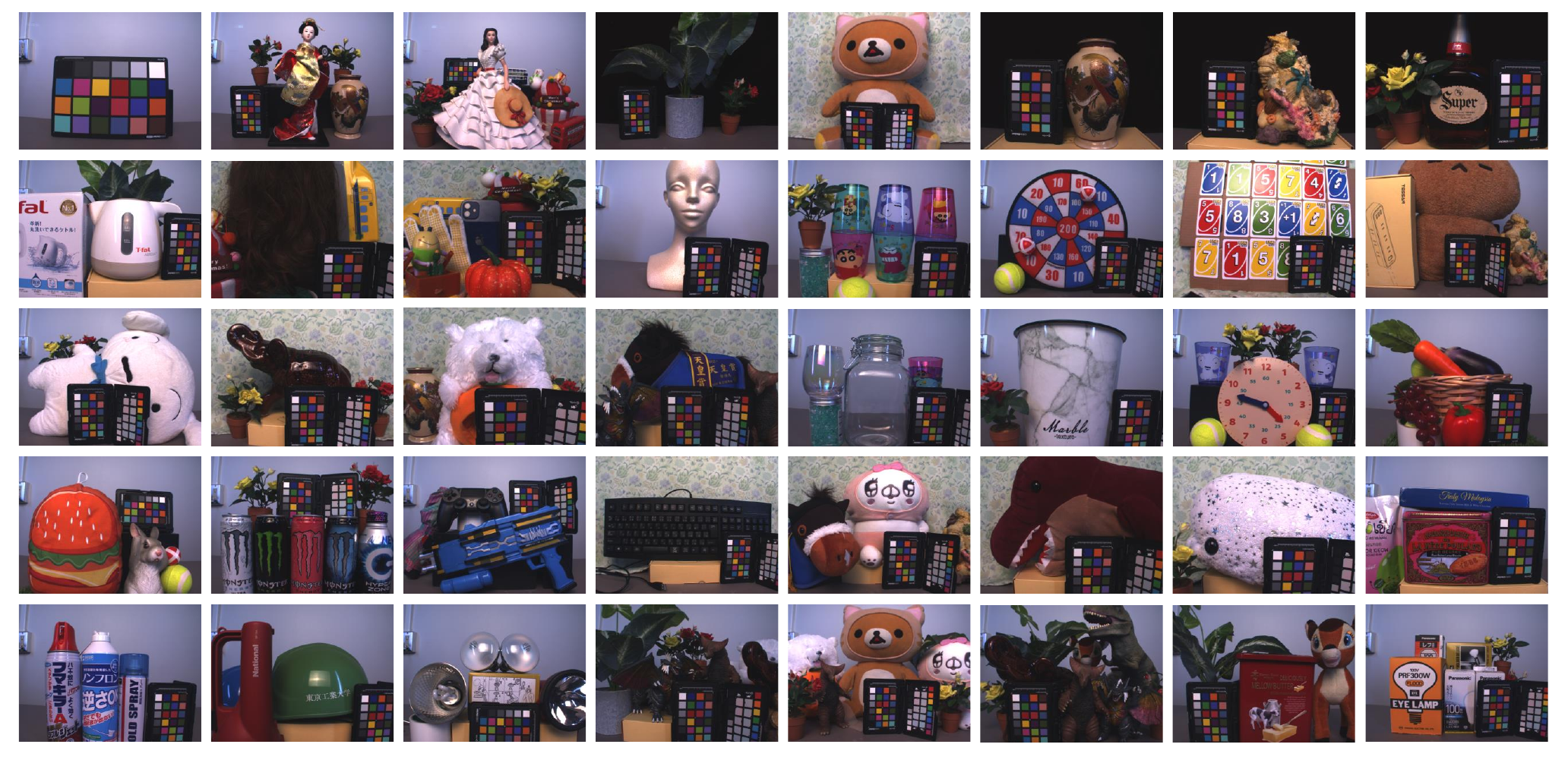}}
\vspace{-2mm}
\caption{40 scenes in our dataset}
\label{fig:datasetfull}
\end{figure}

We constructed a new polarization denoising and demosaicking dataset containing 40 scenes as shown in Fig.~\ref{fig:datasetfull}. Our dataset consists of pairs of ground-truth noise-free full polarization images and corresponding noisy raw mosaic images. The dataset includes three noise-level conditions~(Low, Medium, and High), as summarized in Table~\ref{tab:noiselevel}.

To capture images, we used a JAI CV-M9GE 3-CCD RGB camera~(1024$\times$768 pixels, 10 bits) and placed a SIGMAKOKI SPF-50C-32 linear polarizer attached to a PH-50-ARS rotating polarizer mount in front of the camera. We then captured the images by rotating the polarizer at the angles of 0, 45, 90, and 135 degrees, resulting in full 12-channel data consisting of three color channels and four polarizer angles.

To obtain noise-free full-color ground-truth images, we captured 1,000 noisy RGB images for each capturing at each polarizer angle. Among these 1,000 images, we excluded 100 images to discount potential outliers (e.g., caused by unexpected capturing errors and instability of the capturing). We averaged the rest 900 images to create a ground-truth image that is noise-free approximately. For this process, we calculated the average intensity $\mu_k \ (k=1,2,3,...,1000)$, which is the average pixel value of all color channels and all pixels, for every 1,000 images. Then, we sorted the averaged intensities~$\mu_k$ in ascending and searched the median of them, which is denoted as $\mu_{med}$. Then, we calculated the differences from the median as $|\mu_k-\mu_{med}|$ and excluded 100 images with the largest differences from the median as possible outliers.

\begin{table}[t!]
\centering
\caption{\centering Three noise-level conditions in our dataset}
\vspace{-2mm}
\label{tab:noiselevel}
\resizebox{0.48\textwidth}{!}{ 
\begin{tabular}{|c|c|c|c|c|c|c|}
\hline
\multirow{2}{*}{\shortstack{Noise\\condition}} & \multicolumn{2}{c|}{Gain} & \multirow{2}{*}{\shortstack{Shutter\\speed}} & \multicolumn{3}{c|}{Average noise level} \\
\cline{2-3} \cline{5-7}
 & Analog & Digital & & $\sigma_R$ & $\sigma_G$ & $\sigma_B$ \\
\hline
Low & \( 0\,\mathrm{dB} \) & \(\times2.14\) & \( 1/30\,\mathrm{s}\) & 2.12 & 1.75 & 3.27 \\
Medium & \( 12\,\mathrm{dB} \) & \(\times1.90\) & \( 1/120\,\mathrm{s}\) & 5.16 & 4.29 & 9.08 \\
High & \( 12\,\mathrm{dB} \) & \(\times3.67\) & \( 1/250\,\mathrm{s}\) & 8.62 & 7.31 & 15.79 \\
\hline
\end{tabular}
}
\vspace{1ex}
\end{table}

\begin{table*}[t!]
\caption{\centering Quantitative comparison for MPFA on the high noise-level condition ($\sigma_G = 7.31$)}
\vspace{-2mm}
\label{tab:psnr4}
\centering
\begin{tabular}{|l|cccccccc|c|}
\hline
\multirow{2}{*}{\hspace{13mm} Method} & \multicolumn{8}{c|}{PSNR $\uparrow$} & \ Angle error $\downarrow$\\ \cline{2-10}
& $I_0$ & $I_{45}$ & $I_{90}$ & $I_{135}$ & $S_0$ & $S_1$ & $S_2$ & $DoP$ & $AoP$ \\
\hline 
ICC~\cite{liu2020new} & 32.07 & 31.89 & 32.06 & 31.81 & 34.94 & 36.95 & 36.50 & 23.74 & 43.95 \\
IGRI-2~\cite{morimatsu2021monochrome} & 31.97 & 31.74 & 31.70  & 31.97 & 33.32 & 39.38 & 38.59 & 26.45 & 42.38 \\
LPD~\cite{qiu2021linear} & 31.98 & 31.80  & 31.96 & 31.82 & 37.31 & 33.53 & 33.31 & 20.46 & 41.86 \\
\hline
ICC~\cite{liu2020new} $\rightarrow$ BM3D~\cite{dabov2007image} & 36.69 & 36.19 & 36.67 & 36.10 & 40.42 & 40.67 & 39.69 & 28.97 & 40.71 \\
IGRI-2~\cite{morimatsu2021monochrome} $\rightarrow$ BM3D~\cite{dabov2007image} & 37.41 & 36.70  & 36.68 & 37.37 & 40.36 & 41.96 & 40.61 & 30.13 & 39.72 \\
LPD~\cite{qiu2021linear} $\rightarrow$ BM3D~\cite{dabov2007image} & 34.43 & 34.14 & 34.43 & 34.19 & 40.54 & 35.77 & 35.43 & 23.29 & 39.90  \\
\hline
PFCD $\rightarrow$ IGRI-2 (Ours) & \textbf{40.01} & \textbf{38.90} & \textbf{40.09} & \textbf{38.96} & \textbf{41.18} & \textbf{47.13} & \textbf{43.69} & \textbf{34.31} & \textbf{30.24} \\ 
\hline
\end{tabular}
\vspace{1ex}
\end{table*}

\begin{figure*}[t!]
\centerline{\includegraphics[width=\linewidth]{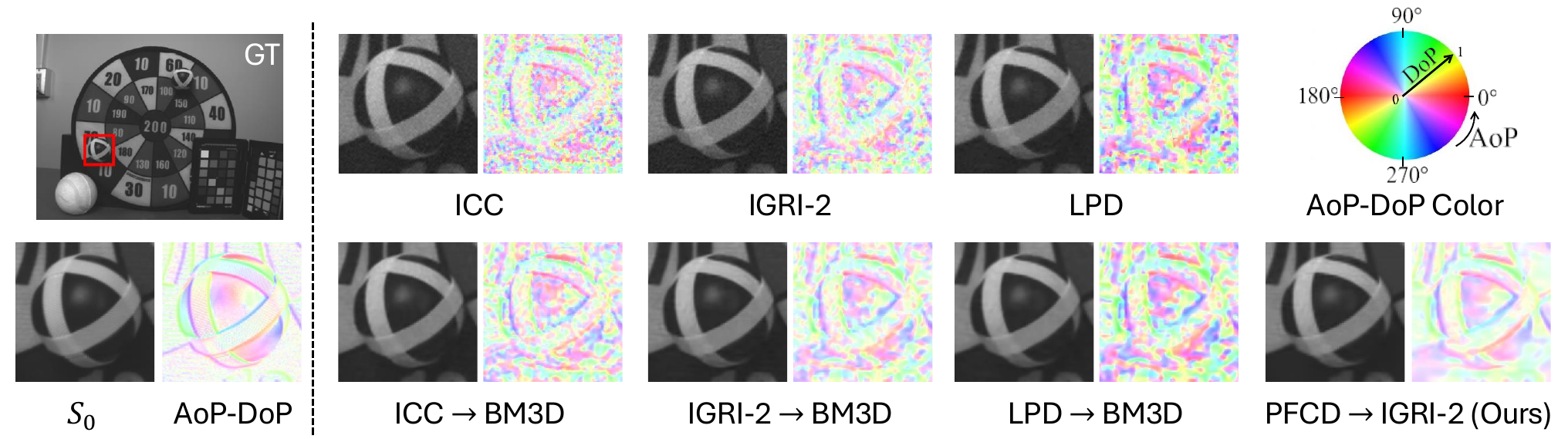}}
\caption{Visual comparison for MPFA on the high noise-level condition (Scene 14)}
\label{fig:ResultMono}
\end{figure*}

To obtain noisy raw mosaic images, we adopted using the noisy RGB images corresponding to the median ($\mu^{\theta}_{med}, \ \theta=\{0,45,90,135\})$ among the 900 images for each polarization angle. We arranged those images to the mosaic image according to the CPFA pattern in Fig.~\ref{fig:OutlineColor}. For the monochrome dataset, we used the green channel and constructed the mosaic image according to the MPFA pattern in Fig.~\ref{fig:OutlineMono}.

The three noise-level conditions in Table~\ref{tab:noiselevel} were adjusted by the combination of the analog gain and the shutter speed. We further applied a digital gain to make the resultant images bright enough for visualization. To determine the digital gains for each condition, we constructed the histogram of all pixel values in 40 scenes and applied the percentile threshold of 99\%, meaning that 99\% of the pixels are ensured unsaturated after multiplying the digital gain. Finally, the noise levels for each condition, polarizer angle, and color channel were calculated by the standard deviation using the before-mentioned 900 noisy RGB images. In real-world data, the noise levels are signal-dependent and vary for every pixel. In Table~\ref{tab:noiselevel}, we provide the average noise levels for three conditions, where the noise levels of all the pixels for all the polarization angles are averaged after applying the digital gains. We observe that the noise levels for the blue channel are relatively higher than the other two channels. This is because of the lower camera spectral sensitivity of the blue channel in the used camera. Table~\ref{tab:noiselevel} shows that our dataset effectively provides images with low, medium, and high noise-level conditions. For simplicity, we use the average noise levels in Table~\ref{tab:noiselevel} for all the scenes and the polarization angles for denoising in later experiments.  

\begin{table*}[t!]
\caption{\centering Quantitative comparison for CPFA on the high noise-level condition ($\sigma_R = 8.62, \sigma_G = 7.31, \sigma_B = 15.79$)}
\vspace{-2mm}
\label{tab:cpsnr5}
\centering
\begin{tabular}{|l|cccccccc|c|}
\hline
\multirow{2}{*}{\hspace{13mm} Method} & \multicolumn{8}{c|}{CPSNR $\uparrow$} & \ Angle error $\downarrow$\\ \cline{2-10}
& $I_0$ & $I_{45}$ & $I_{90}$ & $I_{135}$ & $S_0$ & $S_1$ & $S_2$ & $DoP$ & $AoP$ \\
\hline 
IGRI-2~\cite{morimatsu2021monochrome} & 28.75 & 28.59 & 28.71 & 28.57 & 31.13 & 34.22 & 33.82 & 20.96 & 45.55 \\
LPD~\cite{qiu2021linear} & 29.22 & 28.99 & 29.15 & 29.00 & 31.49 & 33.12 & 32.67 & 19.87 & 41.90 \\
JCPD~\cite{wen2021sparse} & 27.18 & 27.15 & 27.37 & 26.96 & 27.72 & 35.03 & 34.71 & 21.52 & 46.58 \\
\hline
IGRI-2~\cite{morimatsu2021monochrome} $\rightarrow$ BM3D~\cite{dabov2007image} & 31.52 & 31.26 & 31.48 & 31.25 & 34.92 & 35.86 & 35.29 & 23.42 & 44.49 \\
LPD~\cite{qiu2021linear} $\rightarrow$ BM3D~\cite{dabov2007image} & 30.18 & 29.91 & 30.12 & 29.93 & 32.49 & 34.08 & 33.55 & 21.20 & 41.05 \\
JCPD~\cite{wen2021sparse} $\rightarrow$ BM3D~\cite{dabov2007image} & 29.76 & 29.66 & 29.94 & 29.42 & 30.41 & 36.82 & 36.01 & 23.86 & 45.10 \\
\hline
PFCD $\rightarrow$ IGRI-2 (Ours) & \textbf{33.71} & \textbf{33.33} & \textbf{33.72} & \textbf{33.42} & \textbf{35.04} & \textbf{41.12} & \textbf{39.52} & \textbf{29.91} & \textbf{37.97} \\ 
\hline
\end{tabular}
\vspace{1ex}
\end{table*}

\begin{figure*}[t!]
\centerline{\includegraphics[width=\linewidth]{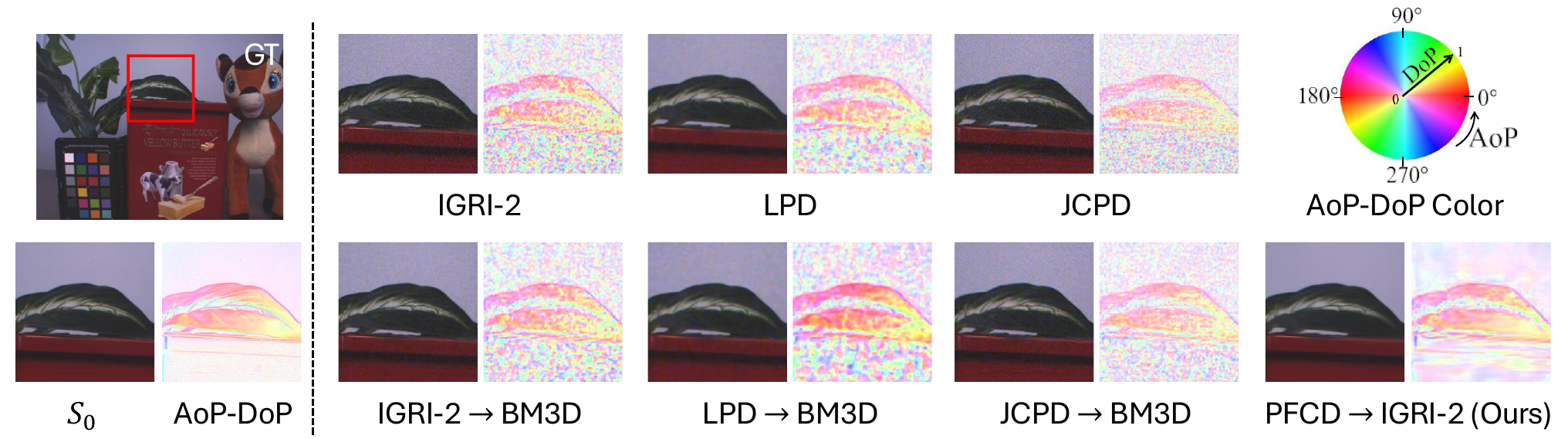}}
\caption{Visual comparison for CPFA on the high noise-level condition (Scene 39)}
\label{fig:ResultColor}
\end{figure*}

\section{Experimental Results}
\label{sec:Experiment}

\subsection{Monochrome Results}
\label{sec:mono}

We first evaluate the results for MPFA. We compared three approaches using publicly available source codes: (i) Demosaicking only; We applied two interpolation-based methods, ICC~\cite{liu2020new} and IGRI-2~\cite{morimatsu2021monochrome}, and one reconstruction-based method, LPD~\cite{qiu2021linear}. (ii) Demosaicking-then-denoising; We applied BM3D denoising~\cite{dabov2007image} for the demosaicked images by each method above, where BM3D was applied in a channel-by-channel manner for each polarization angle image using the average noise level ($\sigma_G$) shown in Table~\ref{tab:noiselevel}. (iii) Denoising-then-demosaicking; We applied our proposed method using the average noise level ($\sigma_G$). 

Table~\ref{tab:psnr4} represents the quantitative comparison on the high noise-level condition, where the average root mean square error (RMSE) of the angle is evaluated for AoP images and the peak signal-to-noise ratio (PSNR) is evaluated for the other images including Stokes images \((S0, S1, S2)\) and DoP images, as in~\cite{morimatsu2021monochrome}. From Table~\ref{tab:psnr4}, we can see that our method outperforms the other methods, especially for DoP and AoP. Figure~\ref{fig:ResultMono} shows the visual comparison, where our method generates the closest result to the ground truth (GT), as can be confirmed in the AoP-DoP visualization.

\vspace{-2mm}
\subsection{Color Results}

We next evaluate the results for CPFA. Similar to the evaluation for MPFA above, we compared three approaches, where we included the reconstruction-based method of JCPD~\cite{wen2021sparse} designed for CPFA, instead of ICC~\cite{liu2020new} designed for MPFA. 

Table~\ref{tab:cpsnr5} represents the quantitative comparison on the high noise-level condition. The results show similar trends to the monochrome case and demonstrate that our method outperforms the other methods in all evaluated categories. Figure~\ref{fig:ResultColor} shows the visual comparison, where only our method can reproduce the AoP-DoP of the leaf region in reasonable quality.

\subsection{Comparison of Three Noise-Level Conditions}

We next evaluate the results of three noise-level conditions. Due to the page limit, we only show the quantitative comparison for MPFA with selected methods and evaluation categories in Table~\ref{tab:psnr_compare}. We refer to the supplementary materials on our website for more complete quantitative and visual results.

In Table~\ref{tab:psnr_compare}, the results for the medium noise level show similar trends to those for the high noise level, where our method achieves the best performance for all the evaluated categories. For the low noise level, IGRI-2$\rightarrow$BM3D and our method show comparable numerical performance, where the best method depends on the evaluation category. This indicates that, for the low noise level, the noise does not much affect the performance of demosaicking, and consequently, demosaicking-then-denoising and denoising-then-demosaicking approaches reach similar performance. Even in this case, our method convincingly shows better AoP-DoP visualization results as shown in the supplementary materials. 

\begin{table}[t!]
\caption{\centering Quantitative comparison for MPFA on three noise-level conditions}
\vspace{-2mm}
\label{tab:psnr_compare}
\centering
\begin{tabular}{|c|l|cc|c|}
\hline
\multirow{2}{*}{\makecell{Noise \\ level}} & \multirow{2}{*}{\hspace{6mm} Method} & \multicolumn{2}{c|}{PSNR $\uparrow$} & \makecell{Angle \\ error} $\downarrow$ \\ \cline{3-5}
& & $S_0$ & $DoP$ & $AoP$ \\ 
\hline 
\multirow{3}{*}{Low} & IGRI-2 & 43.02 & 33.10 & 28.71  \\
& IGRI-2 $\rightarrow$ BM3D & \textbf{45.10} & 33.69 & 25.92  \\
& Ours & 44.92 & \textbf{33.75} & \textbf{21.42}  \\
\hline
\multirow{3}{*}{Medium} & IGRI-2 & 37.42 & 27.94 & 37.56  \\
& IGRI-2 $\rightarrow$ BM3D & 42.54 & 30.73 & 34.48  \\
& Ours & \textbf{42.71} & \textbf{32.63} & \textbf{26.19}  \\
\hline
\multirow{3}{*}{High} & IGRI-2 & 33.32 & 26.45 & 42.38  \\
& IGRI-2 $\rightarrow$ BM3D & 40.36 & 30.13 & 39.72  \\
& Ours & \textbf{41.18} & \textbf{34.31} & \textbf{30.24}  \\ 
\hline
\end{tabular}
\vspace{1ex}
\end{table}

\section{Conclusion}
\label{sec:Conclusion}

In this paper, we have proposed a novel dataset and method for polarization denoising and demosaicking. Our real-world dataset includes a variety of 40 scenes and three noise levels, offering a suitable evaluation platform. Our method effectively utilizes well-accepted signal processing components, offering a solid baseline method. Experimental results on our dataset have demonstrated that our method achieves the best performance both numerically and visually. Our future work includes the consideration of signal-dependent noise for real data and the expansion of data amount for deep-learning-based approaches, which were out of scope in this study.  

\newpage
\ninept
\bibliographystyle{IEEEbib}
\bibliography{strings,refs}

\end{document}